%% file: main.tex
\documentclass[sigconf, 10pt, nonacm]{acmart}
\usepackage[english]{babel}
\usepackage{tikz}
\usepackage{filecontents}

\usepackage{tikz}
\usepackage{xcolor}
\usepackage{xspace}
\usepackage{tikz}
\usepackage{pifont}
\usepackage[english]{babel}
\usepackage{blindtext}
\usepackage{lipsum}

\usepackage[ruled,vlined,linesnumbered]{algorithm2e}
\usepackage{booktabs}

\usepackage{filecontents}
\usepackage{xspace}
\usepackage{xcolor}
\usepackage{hhline}
\usepackage{multirow}
\usepackage{soul}

\usepackage{array}
\usepackage{comment}
\usepackage{listings}
\usepackage{dsfont}

\usepackage{algorithmic}
\usepackage{xurl}
\usepackage{setspace}
\usepackage[flushleft]{threeparttable}
\usepackage{enumerate}
\usepackage{float}
\usepackage{subcaption}
\usepackage{graphicx}
\usepackage{breqn}
\usepackage{enumitem}

\usepackage{caption}
\usepackage{ltablex}

\usepackage{ifluatex}

\usepackage{outlines}

\usepackage{comment}
\usepackage{amsmath, amssymb}
\usepackage{adjustbox}
\input{macros}

\title{An Internet for the KV Cache: Rethinking Classical Infrastructure Boundaries in the LLM Inference Age}

\author{Siddhant Ray, Nick Feamster, Junchen Jiang \\
\emph{The University of Chicago}}

\renewcommand{\shortauthors}{Ray et al.}

\input{abstract}

\begin{document}

\maketitle

\input{intro}

\input{advancements}
\input{kvcache_motivation}

\input{vision}

\input{challenges}
\input{conclusion}

\bibliographystyle{ACM-Reference-Format}
\bibliography{main}

\end{document}

%% file: macros.tex
\definecolor{darkkhaki}{rgb}{0.74, 0.72, 0.42}

\newcounter{packednmbr}
\newenvironment{packedenumerate}{\begin{list}{\thepackednmbr.}{\usecounter{packednmbr}\setlength{\itemsep}{0.5pt}\addtolength{\labelwidth}{-4pt}\setlength{\leftmargin}{2ex}\setlength{\listparindent}{\parindent}\setlength{\parsep}{1pt}\setlength{\topsep}{0pt}}}{\end{list}}
\newenvironment{packeditemize}{\begin{list}{$\bullet$}{\setlength{\itemsep}{0.5pt}\addtolength{\labelwidth}{-4pt}\setlength{\leftmargin}{2ex}\setlength{\listparindent}{\parindent}\setlength{\parsep}{1pt}\setlength{\topsep}{2pt}}}{\end{list}}

\newcommand{\eg}{{\it e.g.,}\xspace}

\newcommand{\mypara}[1]{\vspace{0.05cm}\noindent{\bf {#1}:}~}

\definecolor{backcolour}{rgb}{0.96,0.96,0.96}
\definecolor{codegray}{rgb}{0.5,0.5,0.5}
\definecolor{deepblue}{rgb}{0,0,0.6}
\definecolor{deepred}{rgb}{0.6,0,0}
\definecolor{deepgreen}{rgb}{0,0.5,0}
\lstdefinestyle{mystyle}{
    backgroundcolor=\color{backcolour},   
    commentstyle=\color{codegreen},
    morekeywords={self, True},
    keywordstyle=\color{deepblue},
    numberstyle=\tiny\color{codegray},
    emph={MyClass,__init__,EncodingType,Image},
    emphstyle=\color{deepred},
    stringstyle=\color{deepgreen},
    basicstyle=\ttfamily\footnotesize,
    breakatwhitespace=false,         
    breaklines=true,                 
    captionpos=b,                    
    keepspaces=true,                 
    numbers=left,                    
    numbersep=5pt,                  
    showspaces=false,                
    showstringspaces=false,
    showtabs=false,                  
    tabsize=1
}

\usepackage{tcolorbox}
\tcbuselibrary{listings, skins, breakable, xparse, theorems}
\usepackage{xfrac}

%% file: abstract.tex
\begin{abstract}
LLM inference has become a global-scale, heterogeneous workload spanning agents, retrieval, tool-use, code execution and multi-modal reasoning. These workloads naturally enable context reuse from overlapping inputs, creating a major opportunity to store and reuse the contexts' KV Caches instead of recomputing them. However, model-side advances that shrink the KV Cache and system-side advances that reduce compute, storage, and transfer costs are evolve independently within legacy cloud boundaries. We argue that future inference infrastructure should allow decoupling of compute and KV Cache storage across cloud and datacenters. The network becomes an active distribution channel; bandwidth, latency and pricing directly determines how the KV Cache should be managed. We propose a vision for \emph{an Internet for the KV Cache}, with KV Cache management working as a content-distribution system. In this view, KV Cache storage and recompute decisions are driven by model, infrastructure, and application metrics, to enable adaptive, content-driven decisions for minimizing latency and cost.

\end{abstract}

%% file: intro.tex
\section{Introduction}
\label{sec:intro}

LLM inference has become a dominant workload in the world, with growing heterogeneity in its structure; it now has multi-step agent-based workloads with LLMs, tool calls, code execution and more. A novel opportunity arises here: \emph{KV Cache reuse}. LLM workloads repeatedly revisit overlapping contexts, allowing previously computed inference state of the model input (KV Cache) to be reused instead of being recomputed. The industry has already recognized the storage and reuse of KV Cache as the \emph{most effective tool} to reduce inference cost and delay~\cite{openaiPromptCaching2026,anthropicPromptCaching2025,nvidiaDynamoKVCache2025,googleManagedLustreKVCache2025,vllmPrefixCaching2026,lmcacheKVCache2026,nvidiaDynamoKVCache2025}.

New model-side and system-side optimizations are emerging at a daily level to support both a) cheaper methods to compute the KV Cache and b) cheaper methods to store, manage and reuse the KV Cache. Given the growing scale of LLM inference, we pose the question - \emph{how should the architecture for compute, storage and distribution for KV Cache and LLM inference be designed and evolved over the next decade?}

\mypara{Problem Formulation}The KV Cache is no longer just a storage vs recompute tradeoff: supporting the global scale of LLM inference efficiently has transformed it into an internet-scale content management and delivery problem. Industry and academia are rapidly developing solutions to support an unprecedented scale of KV Cache management~\cite{cheng2025lmcache, qin2025mooncake, bytedanceaibrix2025}.

Currently, the model (LLM) side and system side optimizations for KV Cache are carried out in isolation. Model side optimizations independently aim to reduce the KV Cache computation costs by developing new model architectures, algorithms for efficient attention and harnesses to reduce the memory footprint (more in \S \ref{sec:background}). The KV Cache size becomes smaller while retaining the same representation power. A concrete example: using optimizations like Compressed Sparse Attention, within one generation from Deepseek-V3.2 to Deepseek-V4-Pro, the KV Cache size for 100,000 tokens reduced by $\sim\!\!9\times$, from 9 GB to 1 GB!~\cite{deepseekv4}

Systems are also optimizing for LLM workloads to reduce KV Cache computation costs with a) efficient GPUs to reduce prefill costs and b) newer interconnects to allow faster transfer of KV Caches (more in \S \ref{sec:background}). NVIDIA's Rubin GPU delivers 50 PFLOPS of NVFP4 inference compute~\cite{nvidiaRubin2026}, over the previous generation Blackwell Ultra with 15 PFLOPS~\cite{nvidiaBlackwellUltra2025} - a $3.3\times$ increase. In parallel, storage and transfer bandwidth technology is also becoming more efficient.~\cite{micronHBM42025, micronSSD2025, awsGDSFSx2026, awsP62026} 

Both directions above evolve at different rates, with a common goal of reducing LLM inference costs. Unfortunately, all optimizations are being developed using the legacy connected systems architectures, with tight coupling and co-location of memory and compute, within existing datacenter and cloud boundaries (more in \S \ref{sec:vision}). 

Optimizing for \emph{overall} lower inference cost requires more flexibility in infrastructure rules and movement of KV Caches, which often goes against the assumptions made by the underlying cloud and datacenter systems. With the gains possible from the (re)use of the KV Cache, the definitions of compute, storage and transfer boundaries for large-scale applications, inherited from these systems, may no longer be the correct in the new world of LLM inference.

It now may be cheaper to store the KV Cache (to be reused within the same request) and load it over an ISP internet link, compared to recomputing it. For example, transferring a 1 GB KV Cache over a 10-Gbps Internet link takes about 0.8 seconds (an extremely small fraction of an agentic workload~\cite{abhyankar2026osworld,openaiChatGPTAgent2026}) and costs roughly \$0.09 at AWS’s standard Internet-egress rate~\cite{aws_data_transfer_costs}. For DeepSeek-V4-Pro (1.6T-A49B), this equates to $\sim\!\!100,000$ tokens with its full-precision, compressed cache format. Recomputing the KV Cache on an AWS 8$\times$B200 GPU cluster takes $\sim\!\!15s$ and costs \$0.47~\cite{aws_p6_b200_pricing}, making transferring the 1 GB KV Cache over a fully utilized 10-Gbps link \emph{$\sim\!\!19\times$ faster!} and \emph{$\sim\!\!5.2\times$ cheaper}.

\mypara{Our Vision}Building on emerging trends in LLM inference, we propose a \emph{vision} of an \emph{Internet for the KV Cache}, that places and connects compute and storage nodes beyond cloud and datacenter boundaries. The placement of compute and KV Cache nodes is determined by a function of a) \emph{model-specific metrics}, like prefill FLOPs and KV Cache size; b) \emph{infrastructure-specific metrics}, like GPU cost, storage cost, and network bandwidth; and c) \emph{application-specific metrics}, like reuse frequency, tool-call duration, and number of turns in a request. We envision an adaptive internet infrastructure, where compute and storage placement can be selected and updated using real-time values of these inputs.

Overall, this paper argues that KV Cache reuse has shifted LLM inference from being a compute-storage tradeoff problem to an internet scale content-management problem. In this world, LLM inference acts as an end point and the KV Cache movement is the content distribution system. We call on the community to treat networks and storage as first-class inference resources, and develop new metrics, abstractions, and architectures to optimize \emph{joint} decisions on recompute, store and reuse for internet scale KV Cache management, further pushing the LLM inference efficiency boundaries.

%% file: advancements.tex
\section{Advancements in LLM inference}
\label{sec:background}

LLM inference is advanced by optimizations that reduce the cost of computation, memory, and data transfer. These advances span three knobs: model-side techniques to reduce the  KV Cache computation cost, memory-side techniques to manage KV Cache and context, and system-side techniques to improve scheduling, placement, and transfer efficiency. 

Each knob targets a different bottleneck but exposes the same trend: \emph{inference is no longer limited to GPU computation}. It is a distributed execution problem over compute, memory, storage, and network resources. 
We highlight how they create opportunities for an Internet for the KV Cache.

\subsection{Model (LLM) Side Optimizations}

Recent work has pursued two complementary directions for reducing computation during LLM inference. First, mixture-of-experts architectures reduce per-token computation by activating only selected experts: Mixture of Lookup Experts~\cite{jie2025mole} re-parameterizes experts as lookup tables, Fiddler~\cite{kamahori2025fiddler} coordinates CPU and GPU execution, Klotski~\cite{fang2025klotski} overlaps expert loading with computation through an expert-aware multi-batch pipeline, and K-Transformers~\cite{chen2025ktransformers} provides optimized heterogeneous kernels for CPU-GPU inference. 

Second, sparse-attention methods reduce the amount of historical state that must be recomputed or accessed: Native Sparse Attention\cite{yuan2025nsa} co-designs trainable sparsity with hardware-efficient execution, DuoAttention~\cite{xiao2025duoattention} applies full attention only to retrieval heads, FlexPrefill~\cite{lai2025flexprefill} adapts sparse patterns and computational budgets to each prompt and attention head, and TidalDecode~\cite{yang2025tidaldecode} amortizes token selection across layers during decoding. 


\subsection{KV Cache Optimizations}

Recent work reduces the KV Cache memory footprint along several complementary directions. First, cache blending and editing increases reuse of existing KV Caches opposed to regenerating request-specific copies: CacheBlend~\cite{yao2025cacheblend} selectively recomputes a subset of tokens when fusing cached chunks, Cache-Craft~\cite{agarwal2025cachecraft} repairs reusable RAG chunk caches, EPIC~\cite{hu2025epic} enables position-independent modular cache reuse, and KVEraser~\cite{li2026kveraser} efficiently deletes localized information from KV Caches without recomputing the entire context.

Second, KV Cache compression reduces the size of cached states: TurboQuant~\cite{zandieh2026turboquant} applies online vector quantization, KVzip~\cite{kim2025kvzip} learns query-agnostic eviction through context reconstruction, RocketKV~\cite{behnam2025rocketkv} combines coarse-grained eviction with fine-grained sparse attention, and ChunkKV~\cite{liu2025chunkkv} preserves semantically coherent tokens during eviction. 

Finally, memory compaction and orchestration reduce the context instantiated as KV Cache and manage it efficiently over time. MemAgent~\cite{yu2026memagent} overwrites a bounded memory, Context Folding~\cite{sun2026contextfolding} compresses completed sub-trajectories, HiAgent~\cite{hu2025hiagent} hierarchically summarizes subgoals, and A-Mem~\cite{xu2025amem} links and updates stored experiences. Beyond in-context compaction, ACE~\cite{zhang2026ace} evolves reusable agent playbooks, while RAGCache~\cite{jin2026ragcache}, METIS~\cite{ray2025metis}, and HedraRAG~\cite{hu2025hedrarag} manage retrieved state and generation tradeoffs to control underlying KV Cache states.


\subsection{System Side Optimizations}

System-side optimizations reduce KV Cache transfer overhead and improve GPU utilization along many complementary directions. First, DistServe~\cite{zhong2024distserve} separates prefill and decode execution and independently provisions each phase, Sarathi-Serve~\cite{agrawal2024sarathiserve} introduces chunked prefill and stall-free batching, Mooncake~\cite{qin2025mooncake} extends disaggregation to a multi-tier KV Cache pool, and WindServe~\cite{feng2025windserve} applies stream-based dynamic scheduling to phase-disaggregated inference. 

Second, interconnect-aware transfer systems reduce the cost of moving KV Caches: KVDirect~\cite{chen2025kvdirect} uses tensor-centric GPU-RDMA and pull-based transfers, LMCache~\cite{cheng2025lmcache} batches and pipelines movement across GPU, CPU, storage, and network tiers while supporting NIXL backends, TraCT~\cite{yoon2025tract} replaces NIXL/UCX network transfers with direct GPU--CXL DMA to shared memory, and CXL-SpecKV~\cite{liu2026cxlspeckv} combines CXL-attached FPGA memory with speculative prefetching. 

Finally, scheduling systems adapt execution to workload and hardware conditions: SOLA performs state-aware per-iteration scheduling~\cite{hong2025sola}, ThunderServe~\cite{jiang2025thunderserve} jointly plans placement and rescheduling across heterogeneous GPUs, Seesaw~\cite{su2025seesaw} dynamically re-shards models between inference phases, and Libra~\cite{ruan2026libra} partitions requests into micro-requests to construct SLO-aware batches. 


%% file: kvcache_motivation.tex
\section{The KV Cache is a \emph{first-class citizen} in LLM inference}
\label{sec:kvcache}

\subsection{The KV Cache is not just \emph{another data structure}}

A common view treats the KV Cache as \emph{yet another tensor data structure}, often analogous to model weights in conventional machine learning systems. This view misses its broader role: the KV Cache encodes input knowledge into a unified representation that allows for auto-regressive token generation, conditioned on this knowledge. This representation can be shared and reused across knowledge bases, inputs, and, as new research suggest, even across different models~\cite{liu2026droidspeak}. It is a compact, persistent, steerable representation of contextual knowledge, converted from natural language into an expressive tool. This richness and expressiveness of the KV Cache enables LLMs to efficiently reuse context across environments, workflows, and tasks at global scale.

The KV Cache is becoming a storage, communication, and scheduling primitive for LLM serving, not merely an serving optimization artifact. Several industry solutions and open-source projects such as LMCache~\cite{cheng2025lmcache},TensorRT~\cite{nvidia2026tensorrtllm_kvcache},llm-d~\cite{llmd2026kvcache} and NVIDIA Dynamo~\cite{nvidia2026dynamo} focus on pushing the capabilities of the KV Cache. Popular innovations in this space include  directly by exposing KV Caches across engines and queries, supporting offloading, prefill-decode disaggregation, KV Cache lookup, movement, compression, and orchestration across GPU, CPU, storage, and network layers.

\subsection{(Re)Using the KV Cache allows for \emph{multiple} correct solutions}

The KV Cache introduces a fundamental tradeoff in LLM inference: it can either be recomputed or stored and reused. This tradeoff becomes richer at finer granularity, where systems may combine partial reuse with partial recomputation, or merge individual KV Caches from text segments to reconstruct the KV Cache of logically combined segments~\cite{yao2025cacheblend}. As a result, the answer to whether it is cheaper to recompute or store the KV Cache for future requests is \emph{not universal}. Depending on the underlying infrastructure, the application determined KV Cache reuse pattern and the model architecture, \emph{multiple solutions} can be correct.

We consider the following formulation to understand the storage-recompute tradeoff for the KV Cache. We will look at cost of storage (including transfer cost over the network), calling it the \emph{CacheCost} and the cost of fully recomputation the KV Cache, calling it the \emph{PrefillCost}. We define -

\[
\resizebox{0.97\columnwidth}{!}{$
\begin{aligned}
\mathrm{CacheCost}
&= \text{cost of storing + transferring reusable KV Cache}, \\
\mathrm{PrefillCost}
&= \text{cost of recomputing the prompt prefill}, \\
\mathrm{DeviceCost}_{\mathrm{hr}}
&= \text{storage device cost per hour per MB}, \\
\mathrm{GPUCost}_{\mathrm{hr}}
&= \text{GPU cost per hour}, \\
\mathrm{TransferCost}
&= \text{data transfer cost per MB}, \\
K
&= \text{frequency of KV Cache reuse}, \\
\mathrm{KV}_{\mathrm{MB}}
&= \text{KV cache size in MB}, \\
\mathrm{Prefill}_{\mathrm{TFLOPs}}
&= \text{compute required for prefill in TFLOPs}, \\
\mathrm{PeakFlopRate}
&= \text{peak GPU throughput in TFLOPs/hour}.
\end{aligned}
$}
\]

Using the above, we formulate

\begin{equation}
\label{eq:cache-prefill-cost-ratio}
\begin{adjustbox}{max width=0.92\linewidth}
$
\begin{aligned}
\frac{\mathrm{CacheCost}}{\mathrm{PrefillCost}}
&=
\frac{
\mathrm{DeviceCost}_{\mathrm{hr}} \cdot K \cdot \mathrm{KV}_{\mathrm{MB}}
+
\mathrm{TransferCost} \cdot \mathrm{KV}_{\mathrm{MB}}
}{
\mathrm{GPUCost}_{\mathrm{hr}}
\cdot
\left(
\frac{\mathrm{Prefill}_{\mathrm{TFLOPs}}}
{\mathrm{PeakFlopRate}}
\right)
} \\[0.5em]
&=
\frac{
\left(
\mathrm{DeviceCost}_{\mathrm{hr}} \cdot K
+
\mathrm{TransferCost}
\right)
\cdot
\mathrm{KV}_{\mathrm{MB}}
}{
\mathrm{GPUCost}_{\mathrm{hr}}
\cdot
\left(
\frac{\mathrm{Prefill}_{\mathrm{TFLOPs}}}
{\mathrm{PeakFlopRate}}
\right)
}.
\end{aligned}
$
\end{adjustbox}
\end{equation}

Setting the left hand side ratio $\frac{CacheCost}{PrefillCost}$ (for the break-even condition), we obtain two \emph{new metrics} - 

\begin{equation}
\label{eq:cache-prefill-equality-condition}
\begin{adjustbox}{max width=0.78\linewidth}
$
\begin{aligned}
\frac{
\mathrm{DeviceCost}_{\mathrm{hr}} \cdot K
+
\mathrm{TransferCost}
}{
\mathrm{GPUCost}_{\mathrm{hr}} / \mathrm{PeakFlopRate}
}
&=
\frac{
\mathrm{Prefill}_{\mathrm{TFLOPs}}
}{
\mathrm{KV}_{\mathrm{MB}}
}.
\end{aligned}
$
\end{adjustbox}
\end{equation}

\begin{table*}[t]
\centering
\small
\setlength{\tabcolsep}{4pt}
\begin{tabular}{c l l l r r c}
\toprule
\# & Model & Infrastructure & $K$ 
& Infra-side Ratio & Model-side Ratio & Decision \\
\midrule
1  & DeepSeek V4-Pro   & AWS B200 + S3 Bucket             & 1 day   & 0.436  & 11.410 & Storing cheaper \\
2  & DeepSeek V4-Pro   & AWS B200 + S3 Bucket              & 1 month & 13.083 & 11.410 & Recompute cheaper \\
3  & DeepSeek V4-Pro   & GCP B200 + GCS             & 1 day   & 0.670  & 11.410 & Storing cheaper \\
4  & DeepSeek V4-Pro   & Azure B200 + Blob Hot LRS  & 1 month & 5.512  & 11.410 & Storing cheaper \\
5  & DeepSeek V4-Pro   & GMI B200 + high storage    & 1 day   & 20.250 & 11.410 & Recompute cheaper \\
6  & DeepSeek V4-Flash & AWS B200 + S3 Bucket              & 1 day   & 0.436  & 4.325  & Storing cheaper \\
7  & DeepSeek V4-Flash & AWS B200 + S3 Bucket              & 1 month & 13.083 & 4.325  & Recompute cheaper \\
\bottomrule
\end{tabular}
\caption{Storage versus recomputation decision across models, infrastructure and reuse ratios.}
\label{tab:cache-recompute-decision}
\end{table*}

From Equation \ref{eq:cache-prefill-equality-condition}, we have the term $\frac{\mathrm{Prefill}_{\mathrm{TFLOPs}}}{\mathrm{KV}_{\mathrm{MB}}}$ which represents the \emph{model} side ratio and the term $\frac{\mathrm{DeviceCost}_{\mathrm{hr}} \cdot K + \mathrm{TransferCost}}{\mathrm{GPUCost}_{\mathrm{hr}} / \mathrm{PeakFlopRate}}$ which represents the \emph{infrastructure and application} side ratio. These ratios can be independently calculated. Directly comparing the values of ratios helps choosing between recomputing or storing the KV Cache for optimal cost-saving.

As established in Equations \ref{eq:cache-prefill-cost-ratio},\ref{eq:cache-prefill-equality-condition}, the choice of \emph{any parameter} can change the decision between storing and recomputing the KV Cache, which leads to the claim that \emph{multiple} correct solutions are possible, conditioned on the setup. To demonstrate this further, we show in Table \ref{tab:cache-recompute-decision}, the effect of changes in a \emph{single parameter} changes the decision. Here, we only consider the extremes of full recomputation vs storage, by introducing partial recomputation and KV Cache blending, we can further push the decision boundary of the tradeoff-space.

\subsection{Enabling KV Cache reuse to reduce the \emph{miss-rate} should be the main goal}

With LLM serving becoming a dominant industry workload, prefix caching~\cite{vllm2026prefixcaching} has emerged as a common KV cache management policy due to its simplicity and the natural occurrence of shared prefixes in applications such as multi-turn chat and coding agents. Industry systems find this attractive and therefore heavily optimize for increasing KV Cache \emph{hit-rate}, a metric that prefix caching can improve extensively when requests share the \emph{exact} token prefixes.

However, the hit rate alone is incomplete, as it does not capture the cost of failure. In LLM serving, a cache miss can be disproportionately expensive because it may require recomputing long-context KV Cache. For example, a hit-rate of 90\% is not useful if recomputing the KV Cache in the remaining 10\% cases becomes the computational bottleneck. Prefix caching is brittle to small input changes — even a single token difference will cause a cache miss and prevent KV Cache reuse. This brittleness becomes more limiting as workloads scale and context reuse requirements shift from exact matches to more semantic and approximate matches. 

The distinguishing factor in KV Cache reuse becomes reducing the \emph{miss-rate}. The extent to which a system can push the tradeoff boundary between KV Cache reuse and recomputation is determined by reducing the cost and frequency of KV Cache misses. As recomputation is \emph{quadratic} in the prefix length, smaller improvements in the KV Cache miss-rate compound and lead to significantly greater compute savings.

Enabling non-prefix KV Cache reuse over partially matching contexts is one such way to expand this boundary. This can be achieved by methods such as blending KV Caches of individual RAG chunks or compacted agent memory, in different orders~\cite{yao2025cacheblend,agarwal2025cachecraft}. This allows multiple versions and perturbations of several smaller contexts to reconstruct the overall KV Cache states, making reuse significantly more robust than exact-prefix matching. Reducing the KV Cache miss rate eventually unlocks the true potential which may be realized from KV reuse across requests and workloads.

\subsection{KV Cache innovations pave the way for \emph{next-generation} LLM workloads}

It may be tempting to think that today’s LLM serving ecosystem does not require further optimizations via smarter KV Cache management, especially as many LLM and agent applications may already \emph{appear} good enough. However, LLM inference is already a global-scale service, and both model capabilities and application demands have changed rapidly over the last two years. As adoption expands and the user-base grows, LLM workloads are \emph{very likely} to evolve further from isolated prompts and sessions toward richer, longer-running, and more interconnected applications.

History in computer systems and networks suggests that infrastructure advances often precede the applications that eventually depend on them. Research and engineering on video streaming over HTTP(S), for example, helped make large-scale video delivery practical before video streaming became a dominant web workload~\cite{dobrian2011impact,jiang2012festive,huang2014buffer,yin2015control,ganjam2015c3}. Today, HTTP(S)-based delivery is the foundation for most content-based Internet video streaming applications. Similarly, advancements built around KV Cache management and distribution can provide building blocks for future LLM applications, that need to share knowledge across silos, workflows, and environments. Investing in KV Cache research and infrastructure, can therefore push LLM inference towards lower cost, higher efficiency, and better quality at scale.

%% file: vision.tex
\section{An Internet for the KV Cache}
\label{sec:vision}

\begin{figure*}
    \centering
    \includegraphics[width=0.775\linewidth]{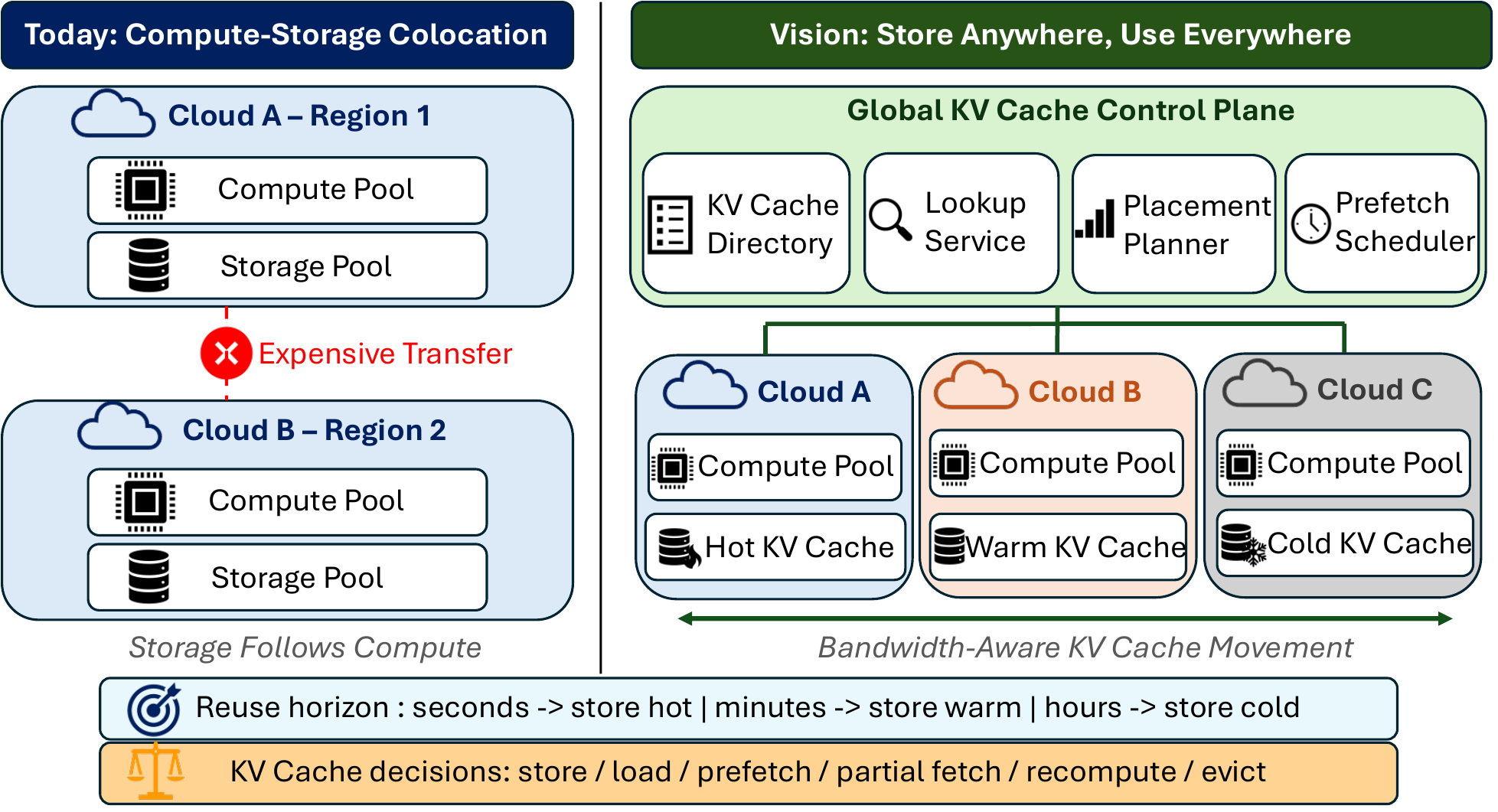}
    \caption{Our Vision: Building a global-scale Internet for KV Cache distribution and management}
    \label{fig:vision}
\end{figure*}

\subsection{How are compute-storage boundaries defined in today's cloud infrastructure?}

Most modern cloud infrastructure is built around the principle that storage and compute should remain close to each other. This is encoded both in architecture and pricing: providers expose regions and availability-zone boundaries as decision knobs for placement. Co-locating storage and compute is recommended to reduce latency and data-transfer costs, and high costs are charged when traffic crosses many of these boundaries. For example, providers like AWS, GCP and Azure use these principles~\cite{aws_s3_perf_guidelines,gcp_storage_locations,azure_files_best_practices}. Data movement across region and zone boundaries has explicit higher pricing~\cite{aws_data_transfer_costs,gcp_vpc_pricing}.

As a result, today’s cloud economics favor tightly coupled placement: keeping data near compute reduces both tail latency and provider-imposed transfer cost. Moving data across zones, regions, or clouds can quickly become a dominant expense. Prior work on cloud transfers reaches the same conclusion, showing that wide-area cloud movement is often slow and that egress prices can dominate the cost of bulk data movement~\cite{jain2022skyplane, liu2025skystore}. Together, these constraints make storage-compute locality not just a performance optimization, but pose core restriction in the way applications deployed on top of cloud systems are priced, and controlled.

Some recent systems have started to push the boundary of \emph{compute across clouds}. SkyPilot introduces an inter-cloud broker that discovers feasible VM instances across zones, regions, and cloud providers, and places jobs based on price, performance, availability, and data-transfer overheads~\cite{yang2023skypilot}. This model is especially attractive for spot-based workloads, where capacity is volatile. SkyServe extends this direction to model serving, by spreading spot and on-demand replicas across regions and clouds and using over-provisioning and on-demand fallback to mask spot unavailability~\cite{mao2025skyserve}.

The above systems make compute placement more flexible, while storage remains a dependency of the execution location. Other work is exploring cross-cloud disaggregated storage at the level of object placement, replication, and bulk transfer~\cite{liu2025skystore}. LLM inference and KV Caches need \emph{more}: they need real-time KV Cache state management with fine-grained placement, prefetching, and recomputation fallback.

\subsection{\emph{Store Anywhere, Use Everywhere} Model}

Our vision (Figure \ref{fig:vision}) is an Internet for the KV Cache which \emph{completely} dissolves cloud boundaries for KV Cache reuse and movement. Current clouds bind storage and compute, making proximity the default metric for efficiency. KV Caches challenge this model. Unlike other user or system data, a KV Cache can be recomputed or reused, based on the time-sensitive inference state. Its optimal location depends not on administrative boundaries, but on the storage-network-recompute tradeoff induced by \emph{future reuse}. Our vision advocates that the Internet for the KV Cache should be fully distributed: it must allow KV Cached state to move across storage tiers, regions, and providers. 

A key insight behind our architecture is that KV Cache reuse is highly heterogeneous along the time domain. In agentic workflows, the same cached context may be reused after seconds, minutes, or even hours, creating reuse horizons that are much longer than the latency of a single generation~\cite{abhyankar2026osworld,openaiChatGPTAgent2026}. This temporal slack enriches the storage-compute tradeoff: KV Caches with near-term reuse can remain close to the GPU, while KV Caches with longer reuse horizons can be pushed to cheaper and farther storage tiers. 

Our vision builds on this principle: the placement of KV Cache and compute should be governed by \emph{jointly} considering reuse rates and deadlines, available bandwidth, transfer and recomputation costs. KV Caches should not be viewed as bound to GPU compute or storage, but as mobile inference states that can be placed, replicated, and delivered wherever needed. We view this as a content-delivery principle: move useful KV Caches toward \emph{expected} future demand, prefetch them when reuse is predictable, and relocate it when demand shifts. Unlike data in traditional CDNs, KV Caches can be regenerated. If delivery becomes too slow or expensive, the system can regenerate the missing KV Caches and re-evaluate placement using updated reuse expectations, freeing the original storage slots for other KV Caches.

For an Internet for the KV Cache, the underlying network is a \emph{first class citizen}: its bandwidth should be used opportunistically to move and distribute KV Caches. This requires a KV Cache control plane with a lookup service that can discover reusable state, estimate transfer deadlines and available bandwidth, and decide when to transfer, replicate or avoid movement of the KV Cache altogether (Figure \ref{fig:vision}). In this world of LLM inference, a global distributed KV Cache Internet, where cached contexts serve as a content distribution system, will ultimately drive LLM performance.

%% file: challenges.tex
\section{Challenges and Opportunities}
\label{sec:challenges}

Building an Internet for the KV Cache requires designing of new communication, policy and data-discovery abstractions . Existing systems provide some useful primitives: prefix caching shows that repeated contexts can be detected and reused; KV offloading shows that inference state can move across memory and storage tiers; prefetching and routing show that workload structure can guide placement decisions; and Internet-style routing shows that large distributed systems can evolve without fixed static paths. An Internet for the KV Cache requires rethinking how compute, storage, and the network are jointly exposed as inference resources.

\mypara{Network-aware KV Cache movement} KV Cache reuse makes the network a first-class part of inference execution. Today, the network is mainly treated as a transport layer between fixed compute and storage locations. In an Internet for the KV Cache, network bandwidth, latency, congestion, and transfer price directly determine whether a KV Cache should be fetched, partially reconstructed, or fully recomputed. 

This may require rethinking interconnect technologies, new transport protocols, and new cache-aware routing mechanisms that can move large KV Cache states predictably across datacenters, cloud regions, edge nodes, and even ISP paths. KV Cache movement should not be hidden behind a static storage abstraction, instead the system should expose data transfer costs and delays as explicit decision inputs.

\mypara{Adaptive infrastructure and control-plane design} Another challenge is to build a connected infrastructure that can adapt as models, workloads, and hardware evolve. An Internet for the KV Cache should provide freedom to add or remove compute and storage nodes and move cached state without binding execution to fixed storage. This requires treating the placement of KV Caches as a dynamic policy decision rather than a static allocation. 

An Internet for the KV Cache needs a control plane which needs to enable distributed lookup, discovery, and routing services for cached inference state. It must find where a KV Cache lives, analyze the costs of bandwidth and transfer deadlines, and select whether to move, replicate, partially fetch, or recompute it. Such a service may build on IP-prefix-style routing primitives but its objective is significantly richer. Routing is no longer about reachability alone, but the lowest-cost path through the storage-network-recompute tradeoff.

The architecture should decide when a KV Cache must be recomputed based on cost and reliability targets, and expose fault-tolerance guarantees. Failures should also be first-class in the design. If transfer is delayed (\eg link failure), or the KV Cache is no longer cost-effective to retrieve, the system should degrade gracefully to full or partial recomputation, and consider falling back to a lower-cost placement policy. The recomputed KV Cache should subsequently be propagated based on the updated placement policy.

Finally, the architecture should jointly expose the independent optimizations emerging in LLM inference as coordinated control knobs rather than isolated mechanisms. Compression, offloading, prefetching, routing, recomputation, scheduling, and model-level reuse should be composed adaptively as workloads evolve.

\mypara{Coordination across regulators, providers, and users} KV Cache placement may now cross administrative and economic boundaries. If KV Cached inference state can be stored far and outside the compute provider, replicated across regions, or moved through third-party networks, we need explicit agreements about pricing, reliability, ownership, privacy, and failure handling and recovery.

These agreements should define who controls the KV Cache, who pays for storage and transfer, who is responsible for correctness, and what happens when retrieval fails or violates a latency target. This is stark contrast from current cloud contracts, where compute, storage, and networking are usually bundled inside provider-defined boundaries. An Internet for the KV Cache will require creation of new coordinating bodies (like the IETF~\cite{ietf}) to standardize transfer, security, and interoperability protocols for KV Cached states.

The Internet for the KV Cache must also allow finer-grained user control. For example, if users can store reusable KV Caches cheaply on their own hardware, there should be support that choice. Privacy and security models must be reconsidered as the infrastructure becomes increasingly decoupled, as both providers and users establish control over compute and storage decisions.

%% file: conclusion.tex
\section{Conclusion}
\label{sec:conclusion}

LLM inference has evolved from a compute problem into a global-scale content-distribution problem centered around KV Cache management. This paper envisions an Internet for the KV Cache, where cached inference state is managed jointly using model, infrastructure, and application metrics, with networks and storage as active knobs for decisions. We position this work as a 
"call for arms", and hope it encourages the community to rethink and explore new directions and research around developing control planes, protocols, abstractions and metrics needed to support an internet-scale KV Cache management system.

%% file: main.bib
@inproceedings{jie2025mole,
  title     = {Mixture of Lookup Experts},
  author    = {Jie, Shibo and Tang, Yehui and Han, Kai and Li, Yitong
               and Tang, Duyu and Deng, Zhi-Hong and Wang, Yunhe},
  booktitle = {Proceedings of the 42nd International Conference on Machine Learning},
  series    = {Proceedings of Machine Learning Research},
  volume    = {267},
  pages     = {27929--27940},
  publisher = {PMLR},
  editor    = {Singh, Aarti and Fazel, Maryam and Hsu, Daniel and Lacoste-Julien, Simon
               and Berkenkamp, Felix and Maharaj, Tegan and Wagstaff, Kiri and Zhu, Jerry},
  month     = {13--19 Jul},
  year      = {2025},
  url       = {https://proceedings.mlr.press/v267/jie25b.html}
}

@inproceedings{kamahori2025fiddler,
  title     = {Fiddler: {CPU--GPU} Orchestration for Fast Inference of
               Mixture-of-Experts Models},
  author    = {Kamahori, Keisuke and Tang, Tian and Gu, Yile
               and Zhu, Kan and Kasikci, Baris},
  booktitle = {The Thirteenth International Conference on Learning Representations},
  year      = {2025},
  url       = {https://proceedings.iclr.cc/paper_files/paper/2025/hash/8cd1ce03ea58b3d7dfd809e4d42f08ea-Abstract-Conference.html}
}

@inproceedings{fang2025klotski,
  title     = {Klotski: Efficient Mixture-of-Expert Inference via
               Expert-Aware Multi-Batch Pipeline},
  author    = {Fang, Zhiyuan and Huang, Yuegui and Hong, Zicong
               and Lyu, Yufeng and Chen, Wuhui and Yu, Yue
               and Yu, Fan and Zheng, Zibin},
  booktitle = {Proceedings of the 30th ACM International Conference on
               Architectural Support for Programming Languages and Operating Systems},
  volume    = {2},
  pages     = {574--588},
  publisher = {Association for Computing Machinery},
  year      = {2025},
  doi       = {10.1145/3676641.3716261},
  url       = {https://doi.org/10.1145/3676641.3716261}
}

@inproceedings{chen2025ktransformers,
  title     = {{KTransformers}: Unleashing the Full Potential of {CPU/GPU}
               Hybrid Inference for {MoE} Models},
  author    = {Chen, Hongtao and Xie, Weiyu and Zhang, Boxin
               and Tang, Jingqi and Wang, Jiahao and Dong, Jianwei
               and Chen, Shaoyuan and Yuan, Ziwei and Lin, Chen
               and Qiu, Chengyu and Zhu, Yuening and Ou, Qingliang
               and Liao, Jiaqi and Chen, Xianglin and Ai, Zhiyuan
               and Wu, Yongwei and Zhang, Mingxing},
  booktitle = {Proceedings of the ACM SIGOPS 31st Symposium on
               Operating Systems Principles},
  pages     = {1014--1029},
  publisher = {Association for Computing Machinery},
  year      = {2025},
  doi       = {10.1145/3731569.3764843},
  url       = {https://doi.org/10.1145/3731569.3764843}
}

@inproceedings{yuan2025nsa,
  title     = {Native Sparse Attention: Hardware-Aligned and Natively
               Trainable Sparse Attention},
  author    = {Yuan, Jingyang and Gao, Huazuo and Dai, Damai
               and Luo, Junyu and Zhao, Liang and Zhang, Zhengyan
               and Xie, Zhenda and Wei, Yuxing and Wang, Lean
               and Xiao, Zhiping and Wang, Yuqing and Ruan, Chong
               and Zhang, Ming and Liang, Wenfeng and Zeng, Wangding},
  booktitle = {Proceedings of the 63rd Annual Meeting of the Association
               for Computational Linguistics (Volume 1: Long Papers)},
  pages     = {23078--23097},
  address   = {Vienna, Austria},
  publisher = {Association for Computational Linguistics},
  editor    = {Che, Wanxiang and Nabende, Joyce and Shutova, Ekaterina
               and Pilehvar, Mohammad Taher},
  month     = jul,
  isbn      = {979-8-89176-251-0},
  year      = {2025},
  doi       = {10.18653/v1/2025.acl-long.1126},
  url       = {https://aclanthology.org/2025.acl-long.1126/}
}

@inproceedings{xiao2025duoattention,
  title     = {{DuoAttention}: Efficient Long-Context {LLM} Inference
               with Retrieval and Streaming Heads},
  author    = {Xiao, Guangxuan and Tang, Jiaming and Zuo, Jingwei
               and Guo, Junxian and Yang, Shang and Tang, Haotian
               and Fu, Yao and Han, Song},
  booktitle = {The Thirteenth International Conference on Learning Representations},
  year      = {2025},
  url       = {https://proceedings.iclr.cc/paper_files/paper/2025/hash/5c1ddd2e59df46fd2aa85c833b1b36ed-Abstract-Conference.html}
}

@inproceedings{lai2025flexprefill,
  title     = {{FlexPrefill}: A Context-Aware Sparse Attention Mechanism
               for Efficient Long-Sequence Inference},
  author    = {Lai, Xunhao and Lu, Jianqiao and Luo, Yao
               and Ma, Yiyuan and Zhou, Xun},
  booktitle = {The Thirteenth International Conference on Learning Representations},
  year      = {2025},
  url       = {https://openreview.net/forum?id=OfjIlbelrT}
}

@inproceedings{yang2025tidaldecode,
  title     = {{TidalDecode}: Fast and Accurate {LLM} Decoding with
               Position Persistent Sparse Attention},
  author    = {Yang, Lijie and Zhang, Zhihao and Chen, Zhuofu
               and Li, Zikun and Jia, Zhihao},
  booktitle = {The Thirteenth International Conference on Learning Representations},
  year      = {2025},
  url       = {https://proceedings.iclr.cc/paper_files/paper/2025/hash/11440c427f0f76f191ac06b50d7a2517-Abstract-Conference.html}
}

@inproceedings{yao2025cacheblend,
  title     = {{CacheBlend}: Fast Large Language Model Serving for {RAG}
               with Cached Knowledge Fusion},
  author    = {Yao, Jiayi and Li, Hanchen and Liu, Yuhan and Ray, Siddhant
               and Cheng, Yihua and Zhang, Qizheng and Du, Kuntai
               and Lu, Shan and Jiang, Junchen},
  booktitle = {Proceedings of the Twentieth European Conference on
               Computer Systems},
  pages     = {94--109},
  publisher = {Association for Computing Machinery},
  year      = {2025},
  doi       = {10.1145/3689031.3696098},
  url       = {https://doi.org/10.1145/3689031.3696098}
}

@article{agarwal2025cachecraft,
  title     = {{Cache-Craft}: Managing Chunk-Caches for Efficient
               Retrieval-Augmented Generation},
  author    = {Agarwal, Shubham and Sundaresan, Sai and Mitra, Subrata
               and Mahapatra, Debabrata and Gupta, Archit
               and Sharma, Rounak and Kapu, Nirmal Joshua
               and Yu, Tong and Saini, Shiv},
  journal   = {Proceedings of the ACM on Management of Data},
  volume    = {3},
  number    = {3},
  pages     = {136:1--136:28},
  publisher = {Association for Computing Machinery},
  year      = {2025},
  doi       = {10.1145/3725273},
  url       = {https://doi.org/10.1145/3725273}
}

@inproceedings{hu2025epic,
  title     = {{EPIC}: Efficient Position-Independent Caching for
               Serving Large Language Models},
  author    = {Hu, Junhao and Huang, Wenrui and Wang, Weidong
               and Wang, Haoyi and Hu, Tiancheng and Qin, Zhang
               and Feng, Hao and Chen, Xusheng and Shan, Yizhou
               and Xie, Tao},
  booktitle = {Proceedings of the 42nd International Conference
               on Machine Learning},
  series    = {Proceedings of Machine Learning Research},
  volume    = {267},
  pages     = {24391--24402},
  publisher = {PMLR},
  editor    = {Singh, Aarti and Fazel, Maryam and Hsu, Daniel and Lacoste-Julien, Simon
               and Berkenkamp, Felix and Maharaj, Tegan and Wagstaff, Kiri and Zhu, Jerry},
  month     = {13--19 Jul},
  year      = {2025},
  url       = {https://proceedings.mlr.press/v267/hu25j.html}
}

@article{li2026kveraser,
  title         = {{KVEraser}: Learning to Steer {KV} Cache for
                   Efficient Localized Context Erasing},
  author        = {Li, Mufei and Liu, Shikun and Fu, Dongqi
                   and Wang, Haoyu and Xia, Yinglong and Li, Hong
                   and Yan, Hong and Li, Pan},
  journal       = {arXiv preprint arXiv:2606.17034},
  month     = jun,
  doi       = {10.48550/arXiv.2606.17034},
  year          = {2026},
  eprint        = {2606.17034},
  archivePrefix = {arXiv},
  primaryClass  = {cs.CL},
  url           = {https://arxiv.org/abs/2606.17034}
}

@inproceedings{zandieh2026turboquant,
  title     = {{TurboQuant}: Online Vector Quantization with
               Near-Optimal Distortion Rate},
  author    = {Zandieh, Amir and Daliri, Majid and Hadian, Majid
               and Mirrokni, Vahab},
  booktitle = {The Fourteenth International Conference on
               Learning Representations},
  year      = {2026},
  url       = {https://openreview.net/forum?id=tO3ASKZlok}
}

@inproceedings{kim2025kvzip,
  title     = {{KVzip}: Query-Agnostic {KV} Cache Compression
               with Context Reconstruction},
  author    = {Kim, Jang-Hyun and Kim, Jinuk and Kwon, Sangwoo
               and Lee, Jae W. and Yun, Sangdoo and Song, Hyun Oh},
  booktitle = {Advances in Neural Information Processing Systems},
  volume    = {38},
  year      = {2025},
  url       = {https://proceedings.neurips.cc/paper_files/paper/2025/hash/f4eaa4b8f2d08edb3f0af990d56134ea-Abstract-Conference.html}
}

@inproceedings{behnam2025rocketkv,
  title     = {{RocketKV}: Accelerating Long-Context {LLM} Inference
               via Two-Stage {KV} Cache Compression},
  author    = {Behnam, Payman and Fu, Yaosheng and Zhao, Ritchie
               and Tsai, Po-An and Yu, Zhiding and Tumanov, Alexey},
  booktitle = {Proceedings of the 42nd International Conference
               on Machine Learning},
  series    = {Proceedings of Machine Learning Research},
  volume    = {267},
  pages     = {3358--3392},
  publisher = {PMLR},
  editor    = {Singh, Aarti and Fazel, Maryam and Hsu, Daniel and Lacoste-Julien, Simon
               and Berkenkamp, Felix and Maharaj, Tegan and Wagstaff, Kiri and Zhu, Jerry},
  month     = {13--19 Jul},
  year      = {2025},
  url       = {https://proceedings.mlr.press/v267/behnam25a.html}
}

@inproceedings{liu2025chunkkv,
  title     = {{ChunkKV}: Semantic-Preserving {KV} Cache Compression
               for Efficient Long-Context {LLM} Inference},
  author    = {Liu, Xiang and Tang, Zhenheng and Dong, Peijie
               and Li, Zeyu and {Liuyue} and Li, Bo
               and Hu, Xuming and Chu, Xiaowen},
  booktitle = {Advances in Neural Information Processing Systems},
  volume    = {38},
  year      = {2025},
  url       = {https://proceedings.neurips.cc/paper_files/paper/2025/hash/2987f911151b39cd3a1761e212319e8e-Abstract-Conference.html}
}

@inproceedings{yu2026memagent,
  title     = {{MemAgent}: Reshaping Long-Context {LLM} with
               Multi-Conv {RL}-Based Memory Agent},
  author    = {Yu, Hongli and Chen, Tinghong and Feng, Jiangtao
               and Chen, Jiangjie and Dai, Weinan and Yu, Qiying
               and Zhang, Ya-Qin and Ma, Wei-Ying and Liu, Jingjing
               and Wang, Mingxuan and Zhou, Hao},
  booktitle = {The Fourteenth International Conference on
               Learning Representations},
  year      = {2026},
  url       = {https://openreview.net/forum?id=k5nIOvYGCL}
}

@inproceedings{sun2026contextfolding,
  title     = {Scaling Long-Horizon Agent via Context Folding},
  author    = {Sun, Weiwei and Lu, Miao and Ling, Zhan and Liu, Kang
               and Yao, Xuesong and Yang, Yiming and Chen, Jiecao},
  booktitle = {Proceedings of the 43rd International Conference
               on Machine Learning},
  year      = {2026},
  note      = {Forthcoming},
  url       = {https://openreview.net/forum?id=JaLXQnA2wi}
}

@inproceedings{hu2025hiagent,
  title     = {{HiAgent}: Hierarchical Working Memory Management
               for Solving Long-Horizon Agent Tasks with
               Large Language Model},
  author    = {Hu, Mengkang and Chen, Tianxing and Chen, Qiguang
               and Mu, Yao and Shao, Wenqi and Luo, Ping},
  booktitle = {Proceedings of the 63rd Annual Meeting of the
               Association for Computational Linguistics
               (Volume 1: Long Papers)},
  pages     = {32779--32798},
  address   = {Vienna, Austria},
  publisher = {Association for Computational Linguistics},
  editor    = {Che, Wanxiang and Nabende, Joyce and Shutova, Ekaterina
               and Pilehvar, Mohammad Taher},
  month     = jul,
  isbn      = {979-8-89176-251-0},
  year      = {2025},
  doi       = {10.18653/v1/2025.acl-long.1575},
  url       = {https://aclanthology.org/2025.acl-long.1575/}
}

@inproceedings{xu2025amem,
  title     = {{A-Mem}: Agentic Memory for {LLM} Agents},
  author    = {Xu, Wujiang and Liang, Zujie and Mei, Kai
               and Gao, Hang and Tan, Juntao and Zhang, Yongfeng},
  booktitle = {Advances in Neural Information Processing Systems},
  volume    = {38},
  year      = {2025},
  url       = {https://proceedings.neurips.cc/paper_files/paper/2025/hash/19909c36f51abc4856b4560aff3d36d6-Abstract-Conference.html}
}

@inproceedings{zhang2026ace,
  title     = {Agentic Context Engineering: Evolving Contexts for
               Self-Improving Language Models},
  author    = {Zhang, Qizheng and Hu, Changran and Upasani, Shubhangi
               and Ma, Boyuan and Hong, Fenglu and Kamanuru, Vamsidhar
               and Rainton, Jay and Wu, Chen and Ji, Mengmeng
               and Li, Hanchen and Thakker, Urmish and Zou, James
               and Olukotun, Kunle},
  booktitle = {The Fourteenth International Conference on
               Learning Representations},
  year      = {2026},
  url       = {https://openreview.net/forum?id=eC4ygDs02R}
}

@article{jin2026ragcache,
  title     = {{RAGCache}: Efficient Knowledge Caching for
               Retrieval-Augmented Generation},
  author    = {Jin, Chao and Zhang, Zili and Jiang, Xuanlin
               and Liu, Fangyue and Liu, Shufan
               and Liu, Xuanzhe and Jin, Xin},
  journal   = {ACM Transactions on Computer Systems},
  volume    = {44},
  number    = {1},
  pages     = {2:1--2:27},
  publisher = {Association for Computing Machinery},
  year      = {2026},
  doi       = {10.1145/3768628},
  url       = {https://doi.org/10.1145/3768628}
}

@inproceedings{ray2025metis,
  title     = {{METIS}: Fast Quality-Aware {RAG} Systems with
               Configuration Adaptation},
  author    = {Ray, Siddhant and Pan, Rui and Gu, Zhuohan
               and Du, Kuntai and Feng, Shaoting
               and Ananthanarayanan, Ganesh and Netravali, Ravi
               and Jiang, Junchen},
  booktitle = {Proceedings of the ACM SIGOPS 31st Symposium on
               Operating Systems Principles},
  pages     = {606--622},
  publisher = {Association for Computing Machinery},
  year      = {2025},
  doi       = {10.1145/3731569.3764855},
  url       = {https://doi.org/10.1145/3731569.3764855}
}

@inproceedings{hu2025hedrarag,
  title     = {{HedraRAG}: Co-Optimizing Generation and Retrieval
               for Heterogeneous {RAG} Workflows},
  author    = {Hu, Zhengding and Murthy, Vibha and Pan, Zaifeng
               and Li, Wanlu and Fang, Xiaoyi and Ding, Yufei
               and Wang, Yuke},
  booktitle = {Proceedings of the ACM SIGOPS 31st Symposium on
               Operating Systems Principles},
  pages     = {623--638},
  publisher = {Association for Computing Machinery},
  year      = {2025},
  doi       = {10.1145/3731569.3764806},
  url       = {https://doi.org/10.1145/3731569.3764806}
}

@inproceedings{zhong2024distserve,
  title     = {{DistServe}: Disaggregating Prefill and Decoding for
               Goodput-Optimized Large Language Model Serving},
  author    = {Zhong, Yinmin and Liu, Shengyu and Chen, Junda
               and Hu, Jianbo and Zhu, Yibo and Liu, Xuanzhe
               and Jin, Xin and Zhang, Hao},
  booktitle = {18th USENIX Symposium on Operating Systems Design
               and Implementation (OSDI 24)},
  pages     = {193--210},
  address   = {Santa Clara, CA},
  publisher = {USENIX Association},
  month     = jul,
  year      = {2024},
  isbn      = {978-1-939133-40-3},
  url       = {https://www.usenix.org/conference/osdi24/presentation/zhong-yinmin}
}

@inproceedings{agrawal2024sarathiserve,
  title     = {Taming Throughput-Latency Tradeoff in {LLM} Inference
               with {Sarathi-Serve}},
  author    = {Agrawal, Amey and Kedia, Nitin and Panwar, Ashish
               and Mohan, Jayashree and Kwatra, Nipun
               and Gulavani, Bhargav and Tumanov, Alexey
               and Ramjee, Ramachandran},
  booktitle = {18th USENIX Symposium on Operating Systems Design
               and Implementation (OSDI 24)},
  pages     = {117--134},
  address   = {Santa Clara, CA},
  publisher = {USENIX Association},
  month     = jul,
  year      = {2024},
  isbn      = {978-1-939133-40-3},
  url       = {https://www.usenix.org/conference/osdi24/presentation/agrawal}
}

@inproceedings{qin2025mooncake,
  title     = {Mooncake: Trading More Storage for Less Computation
               --- A {KVCache-Centric} Architecture for Serving
               {LLM} Chatbot},
  author    = {Qin, Ruoyu and Li, Zheming and He, Weiran
               and Cui, Jialei and Ren, Feng and Zhang, Mingxing
               and Wu, Yongwei and Zheng, Weimin and Xu, Xinran},
  booktitle = {23rd USENIX Conference on File and Storage
               Technologies (FAST 25)},
  pages     = {155--170},
  address   = {Santa Clara, CA},
  publisher = {USENIX Association},
  month     = feb,
  year      = {2025},
  isbn      = {978-1-939133-45-8},
  url       = {https://www.usenix.org/conference/fast25/presentation/qin}
}

@inproceedings{feng2025windserve,
  title     = {{WindServe}: Efficient Phase-Disaggregated {LLM}
               Serving with Stream-Based Dynamic Scheduling},
  author    = {Feng, Jingqi and Huang, Yukai and Zhang, Rui
               and Liang, Sicheng and Yan, Ming and Wu, Jie},
  booktitle = {Proceedings of the 52nd Annual International
               Symposium on Computer Architecture},
  series    = {ISCA '25},
  pages     = {1283--1295},
  publisher = {Association for Computing Machinery},
  year      = {2025},
  doi       = {10.1145/3695053.3730999},
  url       = {https://doi.org/10.1145/3695053.3730999}
}

@article{chen2025kvdirect,
  title         = {{KVDirect}: Distributed Disaggregated {LLM}
                   Inference},
  author        = {Chen, Shiyang and Jiang, Rain and Yu, Dezhi
                   and Xu, Jinlai and Chao, Mengyuan and Meng, Fanlong
                   and Jiang, Chenyu and Xu, Wei and Liu, Hang},
  journal       = {arXiv preprint arXiv:2501.14743},
  doi       = {10.48550/arXiv.2501.14743},
  year          = {2025},
  eprint        = {2501.14743},
  archivePrefix = {arXiv},
  primaryClass  = {cs.DC},
  url           = {https://arxiv.org/abs/2501.14743}
}

@article{cheng2025lmcache,
  title         = {{LMCache}: An Efficient {KV} Cache Layer for
                   Enterprise-Scale {LLM} Inference},
  author        = {Cheng, Yihua and Liu, Yuhan and Yao, Jiayi
                   and An, Yuwei and Chen, Xiaokun and Feng, Shaoting
                   and Huang, Yuyang and Shen, Samuel and Du, Kuntai
                   and Jiang, Junchen},
  journal       = {arXiv preprint arXiv:2510.09665},
  month     = oct,
  doi       = {10.48550/arXiv.2510.09665},
  year          = {2025},
  eprint        = {2510.09665},
  archivePrefix = {arXiv},
  primaryClass  = {cs.DC},
  url           = {https://arxiv.org/abs/2510.09665}
}

@article{yoon2025tract,
  title         = {{TraCT}: Disaggregated {LLM} Serving with
                   {CXL} Shared Memory {KV} Cache at Rack-Scale},
  author        = {Yoon, Dongha and Min, Younghoon and Kim, Hoshik
                   and Noh, Sam H. and Kim, Jongryool},
  journal       = {arXiv preprint arXiv:2512.18194},
  month     = dec,
  doi       = {10.48550/arXiv.2512.18194},
  year          = {2025},
  eprint        = {2512.18194},
  archivePrefix = {arXiv},
  primaryClass  = {cs.DC},
  url           = {https://arxiv.org/abs/2512.18194}
}

@inproceedings{liu2026cxlspeckv,
  title     = {{CXL-SpecKV}: A Disaggregated {FPGA} Speculative
               {KV-Cache} for Datacenter {LLM} Serving},
  author    = {Liu, Dong and Yu, Yanxuan},
  booktitle = {Proceedings of the 2026 ACM/SIGDA International
               Symposium on Field Programmable Gate Arrays},
  series    = {FPGA '26},
  pages     = {56--66},
  publisher = {Association for Computing Machinery},
  year      = {2026},
  doi       = {10.1145/3748173.3779188},
  url       = {https://doi.org/10.1145/3748173.3779188}
}

@inproceedings{hong2025sola,
  title     = {{SOLA}: Optimizing {SLO} Attainment for Large
               Language Model Serving with State-Aware Scheduling},
  author    = {Hong, Ke and Li, Xiuhong and Chen, Lufang and Mao, Qiuli and Dai, Guohao
               and Ning, Xuefei and Yan, Shengen and Liang, Yun and Wang, Yu},
  booktitle = {Proceedings of Machine Learning and Systems},
  volume    = {7},
  year      = {2025},
  url       = {https://proceedings.mlsys.org/paper_files/paper/2025/hash/bc82dbfbfa43232be85b8d9838f49c3e-Abstract-Conference.html}
}

@inproceedings{jiang2025thunderserve,
  title     = {{ThunderServe}: High-Performance and Cost-Efficient
               {LLM} Serving in Cloud Environments},
  author    = {Jiang, Youhe and Fu, Fangcheng and Yao, Xiaozhe
               and Wang, Taiyi and Cui, Bin and Klimovic, Ana
               and Yoneki, Eiko},
  booktitle = {Proceedings of Machine Learning and Systems},
  volume    = {7},
  year      = {2025},
  url       = {https://proceedings.mlsys.org/paper_files/paper/2025/hash/c2a0e26dd9ee7d57e92bb1c24b39659a-Abstract-Conference.html}
}

@inproceedings{su2025seesaw,
  title     = {Seesaw: High-Throughput {LLM} Inference via
               Model Re-Sharding},
  author    = {Su, Qidong and Zhao, Wei and Li, Xin
               and Andoorveedu, Muralidhar and Jiang, Chenhao
               and Zhu, Zhanda and Song, Kevin and Giannoula, Christina
               and Pekhimenko, Gennady},
  booktitle = {Proceedings of Machine Learning and Systems},
  volume    = {7},
  year      = {2025},
  url       = {https://proceedings.mlsys.org/paper_files/paper/2025/hash/cbc4ab80cd77aa0eb87da062fbcddb46-Abstract-Conference.html}
}

@inproceedings{ruan2026libra,
  title     = {Libra: Flexible Request Partitioning and Scheduling
               for Serving Unbalanced and Dynamic {LLM} Workloads},
  author    = {Ruan, Chaoyi and Chen, Yinhe and Tian, Dongqi
               and Shi, Yandong and Wu, Yongji and Li, Jialin
               and Li, Cheng},
  booktitle = {23rd USENIX Symposium on Networked Systems Design
               and Implementation (NSDI 26)},
  pages     = {1243--1258},
  address   = {Renton, WA},
  publisher = {USENIX Association},
  month     = may,
  year      = {2026},
  isbn      = {978-1-939133-54-0},
  url       = {https://www.usenix.org/conference/nsdi26/presentation/ruan-libra}
}

@inproceedings{liu2026droidspeak,
  title = {DroidSpeak: {KV} Cache Sharing Across Fine-tuned Model Variants},
  author = {Liu, Yuhan and Huang, Yuyang and Yao, Jiayi and Feng, Shaoting and Gu, Zhuohan and Du, Kuntai and Li, Hanchen and Cheng, Yihua and Jiang, Junchen and Lu, Shan and Musuvathi, Madan and Choukse, Esha},
  booktitle = {Proceedings of the 23rd USENIX Symposium on Networked Systems Design and Implementation ({NSDI} 26)},
  year = {2026},
  month = may,
  location = {Renton, WA, USA},
  url = {https://www.usenix.org/conference/nsdi26/presentation/liu-yuhan},
  publisher = {USENIX Association}
}

@misc{nvidia2026tensorrtllm_kvcache,
  title        = {{TensorRT-LLM}: {KV} Cache System},
  author       = {{NVIDIA}},
  year         = {2026},
  howpublished = {\url{https://github.com/NVIDIA/TensorRT-LLM/blob/main/docs/source/features/kvcache.md}},
  note         = {Accessed: 2026-07-08}
}

@misc{llmd2026kvcache,
  title        = {{llm-d-kv-cache}: Distributed {KV} Cache Scheduling and Offloading Libraries},
  author       = {{llm-d Contributors}},
  year         = {2026},
  howpublished = {\url{https://github.com/llm-d/llm-d-kv-cache}},
  note         = {Accessed: 2026-07-08}
}

@misc{nvidia2026dynamo,
  title        = {{NVIDIA Dynamo}: A Datacenter-Scale Distributed Inference Framework},
  author       = {{NVIDIA}},
  year         = {2026},
  howpublished = {\url{https://docs.nvidia.com/dynamo/getting-started/introduction}},
  note         = {Accessed: 2026-07-08}
}

@misc{vllm2026prefixcaching,
  title        = {{vLLM}: Automatic Prefix Caching},
  author       = {{vLLM Project}},
  year         = {2026},
  howpublished = {\url{https://docs.vllm.ai/en/stable/design/prefix_caching/}},
  note         = {Accessed: 2026-07-08}
}

@misc{aws_s3_perf_guidelines,
  title        = {Performance Guidelines for {Amazon S3}},
  author       = {{Amazon Web Services}},
  year         = {2026},
  howpublished = {\url{https://docs.aws.amazon.com/AmazonS3/latest/userguide/optimizing-performance-guidelines.html}},
  note         = {Accessed: 2026-07-09}
}

@misc{aws_data_transfer_costs,
  title        = {Overview of Data Transfer Costs for Common Architectures},
  author       = {{Amazon Web Services}},
  year         = {2021},
  month        = jun,
  howpublished = {\url{https://aws.amazon.com/blogs/architecture/overview-of-data-transfer-costs-for-common-architectures/}},
  note         = {Accessed: 2026-07-09}
}

@misc{gcp_storage_locations,
  title        = {Bucket Locations},
  author       = {{Google Cloud}},
  year         = {2026},
  howpublished = {\url{https://cloud.google.com/storage/docs/locations}},
  note         = {Accessed: 2026-07-09}
}

@misc{gcp_vpc_pricing,
  title        = {Network Pricing},
  author       = {{Google Cloud}},
  year         = {2026},
  howpublished = {\url{https://cloud.google.com/vpc/network-pricing}},
  note         = {Accessed: 2026-07-09}
}

@misc{azure_files_best_practices,
  title        = {Architecture Best Practices for {Azure Files}},
  author       = {{Microsoft Azure}},
  year         = {2026},
  howpublished = {\url{https://learn.microsoft.com/en-us/azure/well-architected/service-guides/azure-files}},
  note         = {Accessed: 2026-07-09}
}

@article{jain2022skyplane,
  title         = {{Skyplane}: Optimizing Transfer Cost and Throughput Using Cloud-Aware Overlays},
  author        = {Jain, Paras and Kumar, Sam and Wooders, Sarah and Patil, Shishir G. and Gonzalez, Joseph E. and Stoica, Ion},
  journal       = {arXiv preprint arXiv:2210.07259},
  year          = {2022},
  eprint        = {2210.07259},
  archivePrefix = {arXiv},
  primaryClass  = {cs.DC},
  url           = {https://arxiv.org/abs/2210.07259}
}

@article{liu2025skystore,
  title         = {{SkyStore}: Cost-Optimized Object Storage Across Regions and Clouds},
  author        = {Liu, Shu and Mo, Xiangxi and Hershcovitch, Moshik and Zhang, Henric and Cheng, Audrey and Girmonsky, Guy and Vernik, Gil and Factor, Michael and Bang, Tiemo and Ponnapalli, Soujanya and Crooks, Natacha and Gonzalez, Joseph E. and Harnik, Danny and Stoica, Ion},
  journal       = {arXiv preprint arXiv:2502.20818},
  year          = {2025},
  eprint        = {2502.20818},
  archivePrefix = {arXiv},
  primaryClass  = {cs.DC},
  url           = {https://arxiv.org/abs/2502.20818}
}

@inproceedings{yang2023skypilot,
  title     = {{SkyPilot}: An Intercloud Broker for Sky Computing},
  author    = {Yang, Zongheng and Wu, Zhanghao and Luo, Michael and Chiang, Wei-Lin and Bhardwaj, Romil and Kwon, Woosuk and Zhuang, Siyuan and Luan, Frank Sifei and Mittal, Gautam and Shenker, Scott and Stoica, Ion},
  booktitle = {Proceedings of the 20th USENIX Symposium on Networked Systems Design and Implementation},
  series    = {NSDI '23},
  pages     = {437--455},
  year      = {2023},
  publisher = {USENIX Association},
  url       = {https://www.usenix.org/conference/nsdi23/presentation/yang-zongheng}
}

@inproceedings{mao2025skyserve,
  title     = {{SkyServe}: Serving {AI} Models across Regions and Clouds with Spot Instances},
  author    = {Mao, Ziming and Xia, Tian and Wu, Zhanghao and Chiang, Wei-Lin and Griggs, Tyler and Bhardwaj, Romil and Yang, Zongheng and Shenker, Scott and Stoica, Ion},
  booktitle = {Proceedings of the Twentieth European Conference on Computer Systems},
  series    = {EuroSys '25},
  pages     = {159--175},
  year      = {2025},
  publisher = {Association for Computing Machinery},
  doi       = {10.1145/3689031.3717459},
  url       = {https://doi.org/10.1145/3689031.3717459}
}

@inproceedings{dobrian2011impact,
  title     = {Understanding the Impact of Video Quality on User Engagement},
  author    = {Dobrian, Florin and Sekar, Vyas and Awan, Asad and Stoica, Ion and Joseph, Dilip and Ganjam, Aditya and Zhan, Jibin and Zhang, Hui},
  booktitle = {Proceedings of the ACM SIGCOMM 2011 Conference},
  series    = {SIGCOMM '11},
  pages     = {362--373},
  year      = {2011},
  publisher = {Association for Computing Machinery},
  doi       = {10.1145/2043164.2018478},
  url       = {https://doi.org/10.1145/2043164.2018478}
}

@inproceedings{jiang2012festive,
  title     = {Improving Fairness, Efficiency, and Stability in {HTTP}-Based Adaptive Video Streaming with {FESTIVE}},
  author    = {Jiang, Junchen and Sekar, Vyas and Zhang, Hui},
  booktitle = {Proceedings of the 8th International Conference on Emerging Networking Experiments and Technologies},
  series    = {CoNEXT '12},
  pages     = {97--108},
  year      = {2012},
  publisher = {Association for Computing Machinery},
  doi       = {10.1145/2413176.2413189},
  url       = {https://doi.org/10.1145/2413176.2413189}
}

@inproceedings{huang2014buffer,
  title     = {A Buffer-Based Approach to Rate Adaptation: Evidence from a Large Video Streaming Service},
  author    = {Huang, Te-Yuan and Johari, Ramesh and McKeown, Nick and Trunnell, Matthew and Watson, Mark},
  booktitle = {Proceedings of the 2014 ACM Conference on SIGCOMM},
  series    = {SIGCOMM '14},
  pages     = {187--198},
  year      = {2014},
  publisher = {Association for Computing Machinery},
  doi       = {10.1145/2619239.2626296},
  url       = {https://doi.org/10.1145/2619239.2626296}
}

@inproceedings{yin2015control,
  title     = {A Control-Theoretic Approach for Dynamic Adaptive Video Streaming over {HTTP}},
  author    = {Yin, Xiaoqi and Jindal, Abhishek and Sekar, Vyas and Sinopoli, Bruno},
  booktitle = {Proceedings of the 2015 ACM Conference on Special Interest Group on Data Communication},
  series    = {SIGCOMM '15},
  pages     = {325--338},
  year      = {2015},
  publisher = {Association for Computing Machinery},
  doi       = {10.1145/2785956.2787486},
  url       = {https://doi.org/10.1145/2785956.2787486}
}

@inproceedings{ganjam2015c3,
  title     = {{C3}: {Internet-Scale} Control Plane for Video Quality Optimization},
  author    = {Ganjam, Aditya and Siddiqui, Faisal and Zhan, Jibin and Liu, Xi and Stoica, Ion and Jiang, Junchen and Sekar, Vyas and Zhang, Hui},
  booktitle = {12th USENIX Symposium on Networked Systems Design and Implementation},
  series    = {NSDI '15},
  pages     = {131--144},
  year      = {2015},
  publisher = {USENIX Association},
  address   = {Oakland, CA},
  month     = may,
  isbn      = {978-1-931971-218},
  url       = {https://www.usenix.org/conference/nsdi15/technical-sessions/presentation/ganjam}
}

@misc{deepseekv4,
      title={DeepSeek-V4: Towards Highly Efficient Million-Token Context Intelligence}, 
      author={DeepSeek-AI},
      year={2026},
      eprint={2606.19348},
      archivePrefix={arXiv},
      primaryClass={cs.CL},
      url={https://arxiv.org/abs/2606.19348}, 
}

@misc{nvidiaRubin2026,
  title        = {{NVIDIA} Kicks Off the Next Generation of AI With Rubin},
  author       = {{NVIDIA}},
  year         = {2026},
  month        = jan,
  howpublished = {\url{https://nvidianews.nvidia.com/news/rubin-platform-ai-supercomputer}},
  note         = {NVIDIA Newsroom}
}

@misc{nvidiaBlackwellUltra2025,
  title        = {Inside {NVIDIA} Blackwell Ultra: The Chip Powering the AI Factory Era},
  author       = {Aubrey, Kyle and Stam, Nick},
  year         = {2025},
  month        = aug,
  howpublished = {\url{https://developer.nvidia.com/blog/inside-nvidia-blackwell-ultra-the-chip-powering-the-ai-factory-era/}},
  note         = {NVIDIA Technical Blog}
}

@misc{micronHBM42025,
  title        = {Micron Ships {HBM4} to Key Customers to Power Next-Gen AI Platforms},
  author       = {{Micron Technology}},
  year         = {2025},
  month        = jun,
  howpublished = {\url{https://investors.micron.com/news-releases/news-release-details/micron-ships-hbm4-key-customers-power-next-gen-ai-platforms}},
  note         = {Micron press release}
}

@misc{micronSSD2025,
  title        = {Micron Unveils Portfolio of Industry-First {SSDs} to Power the AI Revolution},
  author       = {{Micron Technology}},
  year         = {2025},
  month        = jul,
  howpublished = {\url{https://investors.micron.com/news-releases/news-release-details/micron-unveils-portfolio-industry-first-ssds-power-ai-revolution}},
  note         = {Micron press release}
}

@misc{awsP62026,
  title        = {Amazon {EC2} {P6e} UltraServers and {P6} Instances},
  author       = {{Amazon Web Services}},
  year         = {2026},
  howpublished = {\url{https://aws.amazon.com/ec2/instance-types/p6/}},
  note         = {AWS product documentation}
}

@misc{awsGDSFSx2026,
  title        = {Accelerate {LLM} Model Loading and Increase Context Windows with {GPUDirect} on Amazon {FSx} for Lustre and TurboQuant},
  author       = {{Amazon Web Services}},
  year         = {2026},
  month        = jun,
  howpublished = {\url{https://aws.amazon.com/blogs/machine-learning/accelerate-llm-model-loading-and-increase-context-windows-with-gpudirect-on-amazon-fsx-for-lustre-and-turboquant/}},
  note         = {AWS Machine Learning Blog}
}

@misc{openaiPromptCaching2026,
  title        = {Prompt Caching 201},
  author       = {{OpenAI}},
  year         = {2026},
  month        = feb,
  howpublished = {\url{https://developers.openai.com/cookbook/examples/prompt_caching_201}},
  note         = {OpenAI Developers Cookbook}
}

@misc{anthropicPromptCaching2025,
  title        = {Prompt Caching with Claude},
  author       = {{Anthropic}},
  year         = {2025},
  month        = aug,
  howpublished = {\url{https://claude.com/blog/prompt-caching}},
  note         = {Claude Blog}
}

@misc{googleManagedLustreKVCache2025,
  title        = {Reducing TCO for AI Inferencing with External KV Cache on Managed Lustre},
  author       = {Shen, Kai and Epstein, Barak},
  year         = {2025},
  month        = oct,
  howpublished = {\url{https://cloud.google.com/blog/products/storage-data-transfer/choosing-google-cloud-managed-lustre-for-your-external-kv-cache}},
  note         = {Google Cloud Blog}
}

@misc{vllmPrefixCaching2026,
  title        = {Automatic Prefix Caching},
  author       = {{vLLM Project}},
  year         = {2026},
  howpublished = {\url{https://docs.vllm.ai/en/latest/features/automatic_prefix_caching/}},
  note         = {vLLM Documentation}
}

@misc{lmcacheKVCache2026,
  title        = {Stop Calling It KV Cache: It’s Something Much Bigger},
  author       = {{LMCache}},
  year         = {2026},
  month        = apr,
  howpublished = {\url{https://blog.lmcache.ai/en/2026/04/28/stop-calling-it-kv-cache-its-something-much-bigger/}},
  note         = {LMCache Blog}
}

@misc{nvidiaDynamoKVCache2025,
  title        = {How to Reduce KV Cache Bottlenecks with NVIDIA Dynamo},
  author       = {Elmeleegy, Amr and Kamath, Akshatha},
  year         = {2025},
  month        = sep,
  howpublished = {\url{https://developer.nvidia.com/blog/how-to-reduce-kv-cache-bottlenecks-with-nvidia-dynamo/}},
  note         = {NVIDIA Technical Blog}
}

@misc{bytedanceaibrix2025,
      title={AIBrix: Towards Scalable, Cost-Effective Large Language Model Inference Infrastructure}, 
      author={The AIBrix Team and Jiaxin Shan and Varun Gupta and Le Xu and Haiyang Shi and Jingyuan Zhang and Ning Wang and Linhui Xu and Rong Kang and Tongping Liu and Yifei Zhang and Yiqing Zhu and Shuowei Jin and Gangmuk Lim and Binbin Chen and Zuzhi Chen and Xiao Liu and Xin Chen and Kante Yin and Chak-Pong Chung and Chenyu Jiang and Yicheng Lu and Jianjun Chen and Caixue Lin and Wu Xiang and Rui Shi and Liguang Xie},
      year={2025},
      eprint={2504.03648},
      archivePrefix={arXiv},
      primaryClass={cs.DC},
      url={https://arxiv.org/abs/2504.03648}, 
}

@misc{abhyankar2026osworld,
      title={OSWorld-Human: Benchmarking the Efficiency of Computer-Use Agents}, 
      author={Reyna Abhyankar and Qi Qi and Yiying Zhang},
      year={2026},
      eprint={2506.16042},
      archivePrefix={arXiv},
      primaryClass={cs.AI},
      url={https://arxiv.org/abs/2506.16042}, 
}

@misc{openaiChatGPTAgent2026,
  title        = {{ChatGPT} Agent},
  author       = {{OpenAI}},
  year         = {2026},
  howpublished = {\url{https://help.openai.com/en/articles/11752874-chatgpt-agent}},
  note         = {OpenAI Help Center}
}
